\documentclass[aip,amsmath,amssymb,reprint]{revtex4-1}  

\usepackage{graphicx}  
\usepackage{dcolumn}   
\usepackage{bm}        
\usepackage{amssymb}   
\graphicspath{{Pictures/}} 
\usepackage{booktabs}
\usepackage{graphicx}
\usepackage{epsfig,amsmath,amssymb,color,float,setspace}
\usepackage[utf8]{inputenc}
\usepackage[T1]{fontenc}
\usepackage{titlesec}
\usepackage{booktabs}
\usepackage{booktabs}
\usepackage[bottom]{footmisc}
\usepackage{multirow}

\usepackage{hhline}

\begin{document}

\preprint{AIP/123-QED}

\title{Exchange-correlation corrections for electronic properties of  \\  half-metallic Co$_2$FeSi and nonmagnetic semiconductor CoFeTiAl}

\author{Olga N. Miroshkina}\email[Author to whom correspondence should be addressed: ]{miroshkina.on@yandex.ru}
\affiliation{Faculty of Physics, Chelyabinsk State University, 454001 Chelyabinsk, Russia}
\affiliation{School of Engineering Science, LUT University, 53850 Lappeenranta, Finland}
\author{Danil R. Baigutlin}
\affiliation{Faculty of Physics, Chelyabinsk State University, 454001 Chelyabinsk, Russia}
\affiliation{School of Engineering Science, LUT University, 53850 Lappeenranta, Finland}
\author{Vladimir V. Sokolovskiy}
\affiliation{Faculty of Physics, Chelyabinsk State University, 454001 Chelyabinsk, Russia}
\affiliation{National University of Science and Technology "MISiS", 119049 Moscow, Russia}
\author{Mikhail A. Zagrebin}
\affiliation{Faculty of Physics, Chelyabinsk State University, 454001 Chelyabinsk, Russia}
\affiliation{National Research South Ural State University, 454080 Chelyabinsk, Russia}
\author{Aki Pulkkinen}\email[Author to whom correspondence should be addressed: ]{aki.pulkkinen@lut.fi}
\affiliation{School of Engineering Science, LUT University, 53850 Lappeenranta, Finland}

\author{Bernardo Barbiellini}
\affiliation{School of Engineering Science, LUT University, 53850 Lappeenranta, Finland}
\affiliation{Physics Department, Northeastern University, Boston, MA 02115, USA}
\author{Erkki L\"ahderanta}
\affiliation{School of Engineering Science, LUT University, 53850 Lappeenranta, Finland}
\author{Vasiliy~D.~Buchelnikov}
\affiliation{Faculty of Physics, Chelyabinsk State University, 454001 Chelyabinsk, Russia}
\affiliation{National University of Science and Technology "MISiS", 119049 Moscow, Russia}

\date{\today}

\begin{abstract}
We consider two cobalt-based full-Heusler compounds CoFeTiAl and Co$_2$FeSi, 
for which Coulomb correlation effects play an important role.
Since the standard GGA scheme does not provide a precise description of the electronic properties near the Fermi level,  we use a meta-GGA functional capable to improve the description of the electronic properties of CoFeTiAl and Co$_2$FeSi. 
In particular, we find a better agreement with the experiment for the magnetic moment and the energy-band gap. Moreover, our calculations show that pressure enhances the insulating properties of Co$_2$FeSi and CoTiFeAl.

\end{abstract}

\pacs{}
\maketitle

\section{Introduction}
Co$_2$-based Heusler compounds are among the best half-metal ferromagnets~(HMF) ~\cite{Felserbook2013}. These alloys exhibit 100\% spin polarization of the electronic density of states~(DOS) at the Fermi energy  ($E_F$) because the band gap is present only for one spin-resolved band.
Moreover, Co$_2$-based Heusler alloys have large total magnetic moments ($\mu_{tot}$) and high Curie temperatures ($T_C$). 
Therefore, they are promising candidates for spintronic applications such as tunnel spin junction and storage devices. 

Among the ternary Co$_2$-based systems, Co$_2$FeSi alloy has been extensively studied both experimentally~\cite{Whurmel_2005,Wurmehl_2006, Deka-2014,Balkeapl2007,Umetsu2012} and theoretically.~\cite{Whurmel_2005,Comtesse20015, Zagrebin_2016,Meinert_2012_prb}
According to experiments, HMF Co$_2$FeSi crystallizes in the L2$_1$ ordered structure (space group $Fm\bar3m$) with a lattice parameter $a_0$ of $5.64$~\AA~\cite{Whurmel_2005, Balkeapl2007} and magnetic moment $\mu_{tot}$ of 6~$\mu_B$/f.u., while the Curie temperature is about 1100~K.~\cite{Whurmel_2005, Deka-2014,Umetsu2012}.

Regarding first-principles studies, the generalized gradient approximation (GGA) gives a lattice parameter $a_0=5.62$~\AA~\cite{ Zagrebin_2016} in agreement with experiments. However, GGA does not reproduce the observed half-metallic behavior  and the total magnetic moment $\mu_{tot}=5.5\ \mu_B$/f.u.~\cite{Comtesse20015, Zagrebin_2016,Meinert_2012_prb} contradicts the Slater-Pauling rule~\cite{Slater-Pauling}.
Correcting GGA with a Hubbard $U$ parameter results in an integer magnetic moment of $6~\mu_B$/f.u. and half-metallic behavior.~\cite{Balke-2006,Whurmel_2005,Zagrebin_2016}.

$GW$ many-body perturbation theory provides another way to correct GGA without adjustable parameters. This approach improves the agreement with the experimental quasiparticle spectra and yields  $\mu_{tot}=5.89$~$\mu_B$/f.u., which is closer to the integer experimental value. \cite{Meinert_2012_prb}.

The quaternary Co-based Heusler alloys belong to another family of half metal compounds.~\cite{Ozdogan_2013} 
They have a generic formula  (XX$^{\prime}$)YZ, in which X, X$^{\prime}$, and Y are transition metals, while Z is an \textit{sp} element.
A member of this family is the non-magnetic semiconductor CoFeTiAl~\cite{Basit} 
possessing an inverse Heusler structure (space group $F4\bar{3}m$) with $a= 5.8509$~\AA.
The GGA electronic structure calculations predict an almost-gapless semiconductor, while the $GW$ increases the energy band gap. Therefore, CoFeTiAl becomes a narrow-band semi-conductor~\cite{Tas} within $GW$. 

Clearly, corrections beyond GGA play an important role in electronic structure of Co$_2$FeSi and CoFeTiAl. The $GW$ scheme is a step in the good direction but the method requires a large  computational effort. Another way to include corrections beyond GGA with less computational effort is given by the meta-GGA strongly constrained and appropriately normed (SCAN)  scheme~\cite{ Sun-2015}. 
A recent Theorem by Perdew {\em et al.} \cite{perdew2017}  on the generalized Kohn–Sham theory explains why SCAN outperforms GGA for energy-band gaps. 
Perdew’s theorem also explains why Generalized Kohn-Sham band gaps from hybrid functionals such as SCAN0 \cite{hui2016} can be more realistic than those from GGAs or even from the exact Kohn-Sham  potential. SCAN0 has shown to perform better than SCAN in transition metal chemistry \cite{modrzejewski2018}. However, since hybrid functionals present problems in metals \cite{paier2007}, SCAN could be a safer choice for Heusler alloys.

Here, we study structural, magnetic, and electronic properties of Co-based full-Heusler compounds Co$_2$FeSi and CoFeTiAl by comparing results obtained with GGA and SCAN. We also study how the energy gap evolves under pressure.

\begin{table*}[!htb] 
    \caption{ The equilibrium lattice parameter~$a_0$,  total magnetic moment $\mu_{tot}$, and the transition energies between certain high-symmetry points for  Co$_2$FeSi and CoFeTiAl calculated with GGA and SCAN in comparison with the available experimental data. $N_V$ is the number of valence electrons. }
    \begin{ruledtabular}
    \begin{tabular}{lcccccccccccccc}
        & \multirow{2}{*}{$N_V$} & \multicolumn{3}{c}{$a_0$ [\AA]} & \multicolumn{3}{c}{$\mu_{tot}$ [$\mu_B$/f.u.]} & \multicolumn{2}{c}{$E_{\Gamma-\Gamma}$ [eV]} &
        \multicolumn{2}{c}{$E_{X-X}$ [eV]} &
        \multicolumn{2}{c}{$E_{\Gamma-X}$ [eV]}\\ 
        \cline{3-5} \cline{6-8} \cline{9-10} \cline{11-12} \cline{13-14} 
         & & GGA & SCAN & exp. & GGA & SCAN & exp. & GGA & SCAN & GGA & SCAN & GGA & SCAN \\
        \hline
        Co$_2$FeSi & 30 & 5.625 & 5.570 & 5.64~$^{\mathrm{Ref.\cite{Balke-2007}}}$ & 5.534 & 6.02 & (5.97$\pm$0.05)$^{\mathrm{Ref.\cite{Balke-2006}} }$ & 0.773 & 2.139 & -- & -- & -- & --\\  
        CoFeTiAl & 24 & 5.806 & 5.756 &  5.82 $^{\mathrm{Ref.\cite{Lin-2016}}}$ & 0 & 0 & 0$^{\mathrm{Ref.\cite{Felserbook2013}}}$ & 0.029 & 0.547 & 0.389 & 0.910 & 0.057 & 0.607
    \end{tabular}
    \end{ruledtabular}
    \label{table1}
\end{table*}

\section{Computational Details}
Density Functional Theory~(DFT) within the projector augmented wave~(PAW) method implemented in VASP code~\cite{Kresse-1996,paw} was used for structural and electronic properties calculations. 
The exchange-correlation effects were included within  GGA ~\cite{Perdew-1991,Burke-1997} and also beyond. There are several GGA functionals \cite{barbiellini1990,lehtola2018}, but in our study, GGA refers to the functional parametrized by Perdew, Burke and Ernzerhof~\cite {Perdew1996} and all our GGA computations use this scheme. Our meta-GGA corrections in the present study are within the SCAN method~\cite{ Sun-2015}. 
The four-atom cells with space groups $Fm\overline{3}m$ and $F\overline{4}3m$ were used for Co$_2$FeSi and CoFeTiAl, respectively.
For ternary compound, two Co atoms occupy the 8\textit{c} (1/4,
1/4, 1/4) and (3/4, 3/4, 3/4) Wyckoff
positions, while Fe and Si occupy the 4\textit{b} (1/2, 1/2, 1/2) and 4\textit{a} (0, 0, 0).
For quaternary alloy, Co, Fe, Ti, and Al occupy 4\textit{c}~(1/4, 1/4, 1/4), 4\textit{d}~(3/4, 3/4, 3/4), 4\textit{b}~(1/2, 1/2, 1/2), and 4\textit{a}~(0, 0, 0) Wyckoff positions, respectively. 
Ferromagnetic  ordering was considered.  

For both compositions, full geometry optimization was performed followed by accurate bands and densities of states (DOS) calculations. 
Convergence threshold for the ionic relaxation was set to 10$^{-3}$~eV/\AA.
The kinetic-energy cutoffs for the plane-wave basis and augmentation charges were 700~eV and 800~eV, respectively.
The Brillouin zone integration was performed using the second-order Methfessel-Paxton method with smearing width of 0.2 eV.
A Monkhorst-Pack $ 12 \times 12 \times 12 $ \textit{k}-points mesh was used for ionic relaxation. A more accurate $ 24 \times 24 \times 24 $ \textit{k}-points mesh was used for bands and DOS calculations.

\begin{figure}[!t]
    \centerline{\includegraphics[width=9.5cm,clip]{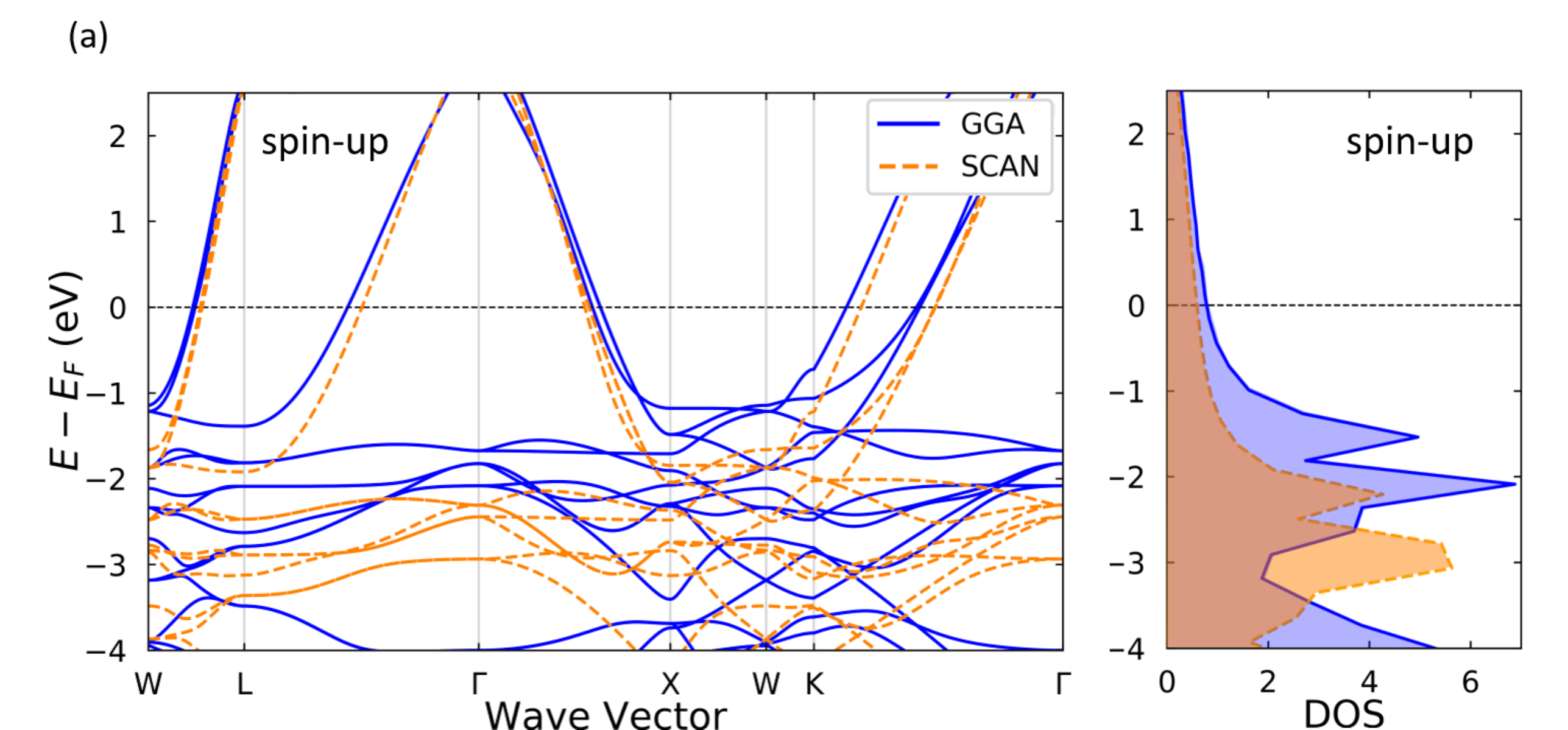}}
    \vfill 
    \centerline{\includegraphics[width=9.5cm,clip]{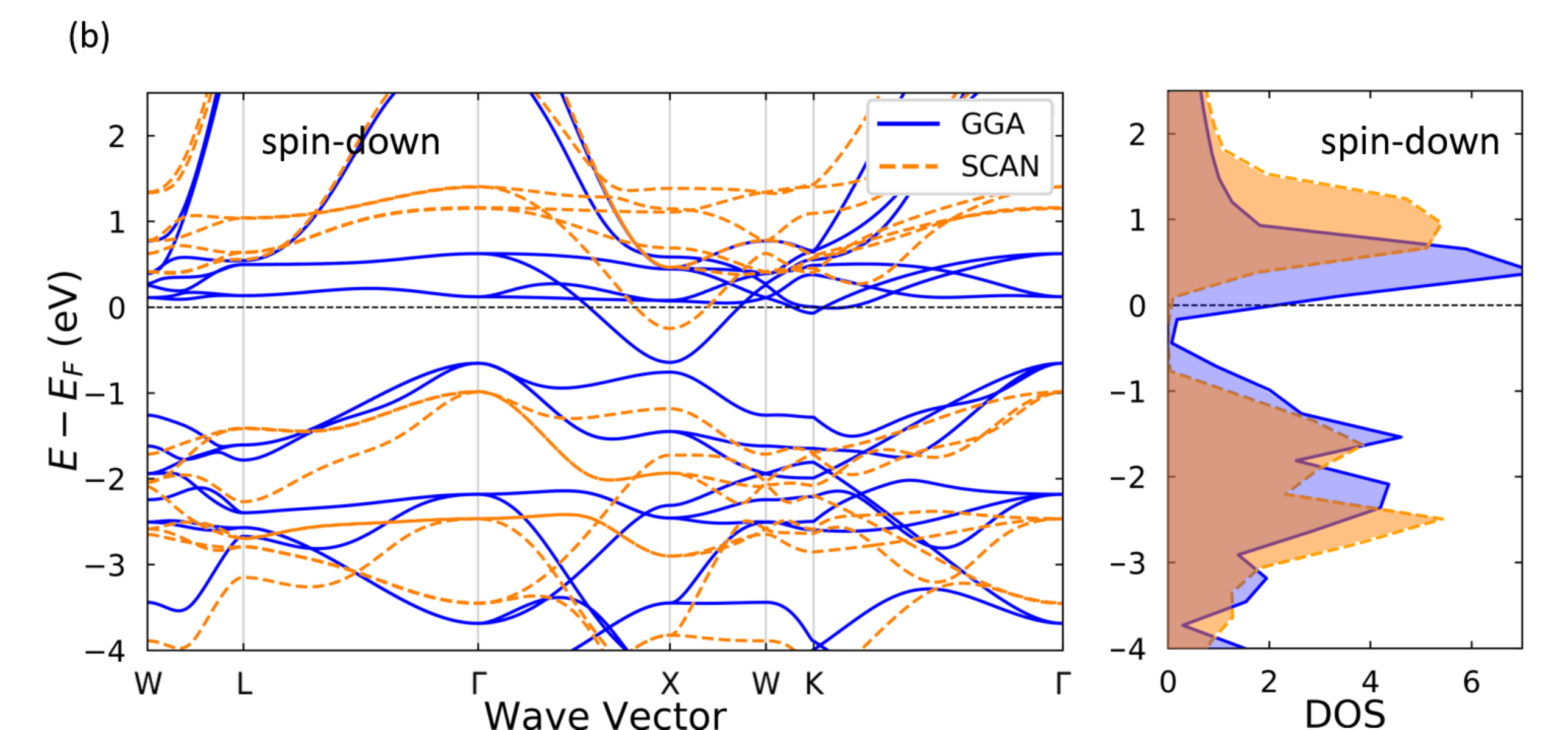}     }
    \caption{Band structure and total DOS of (a)~majority and (b)~minority spin band for Co$_2$FeSi  calculated with GGA and SCAN.}
    \label{fig1}
\end{figure}

\section{Results and Discussion}
The results of structural and magnetic properties calculations are summarized in Table~\ref{table1}.
SCAN yields a smaller lattice parameter in comparison with GGA.
GGA and $GW$~\cite{Meinert_2012_prb} give non-integer values of the magnetic moment, thus SCAN predicts integer $\mu_{tot}$ in agreement with experiments. 

Figure~\ref{fig1} illustrates the energy-band structures and total DOS for the minority and majority channels for Co$_2$FeSi.
The GGA and SCAN DOS for the majority spin band (Fig.~\ref{fig1}(a)) have the occupied band at the Fermi level~($E_F$) showing a metallic character. 
Regarding the minority spin band (Fig.~\ref{fig1}(b)), GGA predicts a half-metallic-like behavior due to the pseudo-gap $E_g$ shifted slightly to lower energies with respect to $E_F$.
However, SCAN produces an exchange splitting $\approx 0.9$~eV of the majority and minority spin states.
Therefore, the SCAN DOS for the minority spin reveals an energy gap 0.91~eV near $E_F$. In the spin-down band structure at the $X$ point, the SCAN conduction band extends significantly less below $E_F$ than the corresponding GGA band.
Although GGA and SCAN give qualitatively different behavior for band structure between all considered high symmetry points in the Brillouin zone of Co$_2$FeSi, both suggest almost half-metallic behavior.
The energies in the valence and conductive bands push away from the Fermi level for both spin channels indicating the presence of exchange splitting.

\begin{figure}[!t]
    \centerline{
    \includegraphics[width=9.5cm,clip]{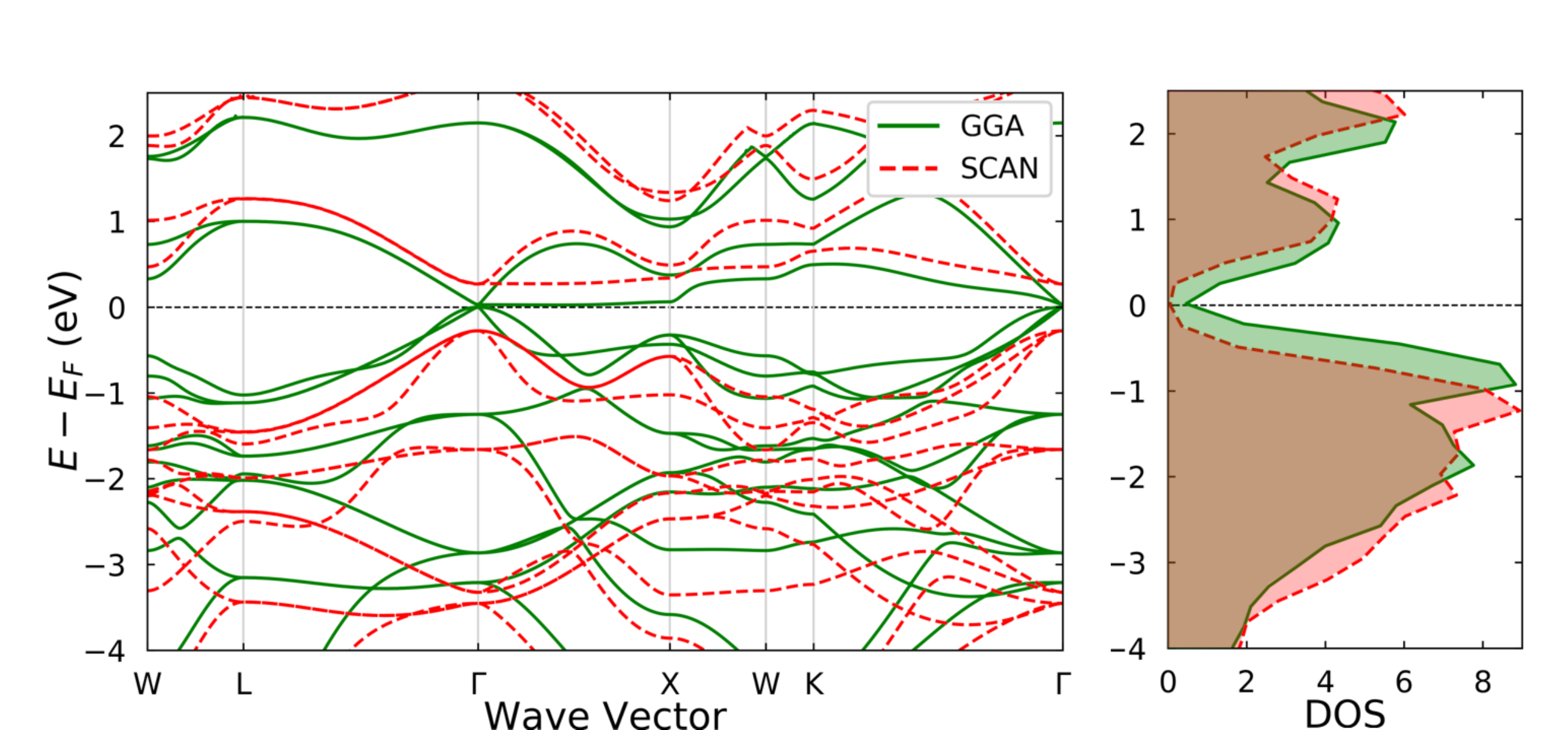}
    }
    \caption{Electronic band structure and total DOS for CoFeTiAl calculated with GGA and SCAN.}
    \label{fig2}
\end{figure}

We now consider the electronic structure of non-magnetic CoFeTiAl presented in Fig.~\ref{fig2}.
For GGA, we obtain an almost-gapless semiconductor behavior with a pseudogap of $\approx 0.03$~eV as in previous studies,~\cite{Tas} while SCAN presents more semiconducting character by separating the valence and the conduction bands. The SCAN band gap is 0.55~eV  characteristic of a semiconductor. The $GW$ result gives a narrow-band semiconductor with $E_{g}^{GW}=0.30$~eV.~\cite{Tas}
A similar situation occurs for  the nonmagnetic semimetal Fe$_2$VAl  according to Buchelnikov {\em et al.}~\cite{Buchelnikov-2019}
The direct $\Gamma-\Gamma$ and $X-X$ and indirect $\Gamma-X$ energy gaps are summarized in Table~\ref{table1}. 
One can see that the $\Gamma-\Gamma$ energy gap is the smallest one, therefore, one can infer conduction properties by examining this gap. 
SCAN energy gaps between high symmetry points are several times larger than GGA values.

\begin{figure}[!hbt]
    \centerline{\includegraphics[width=8cm,clip]{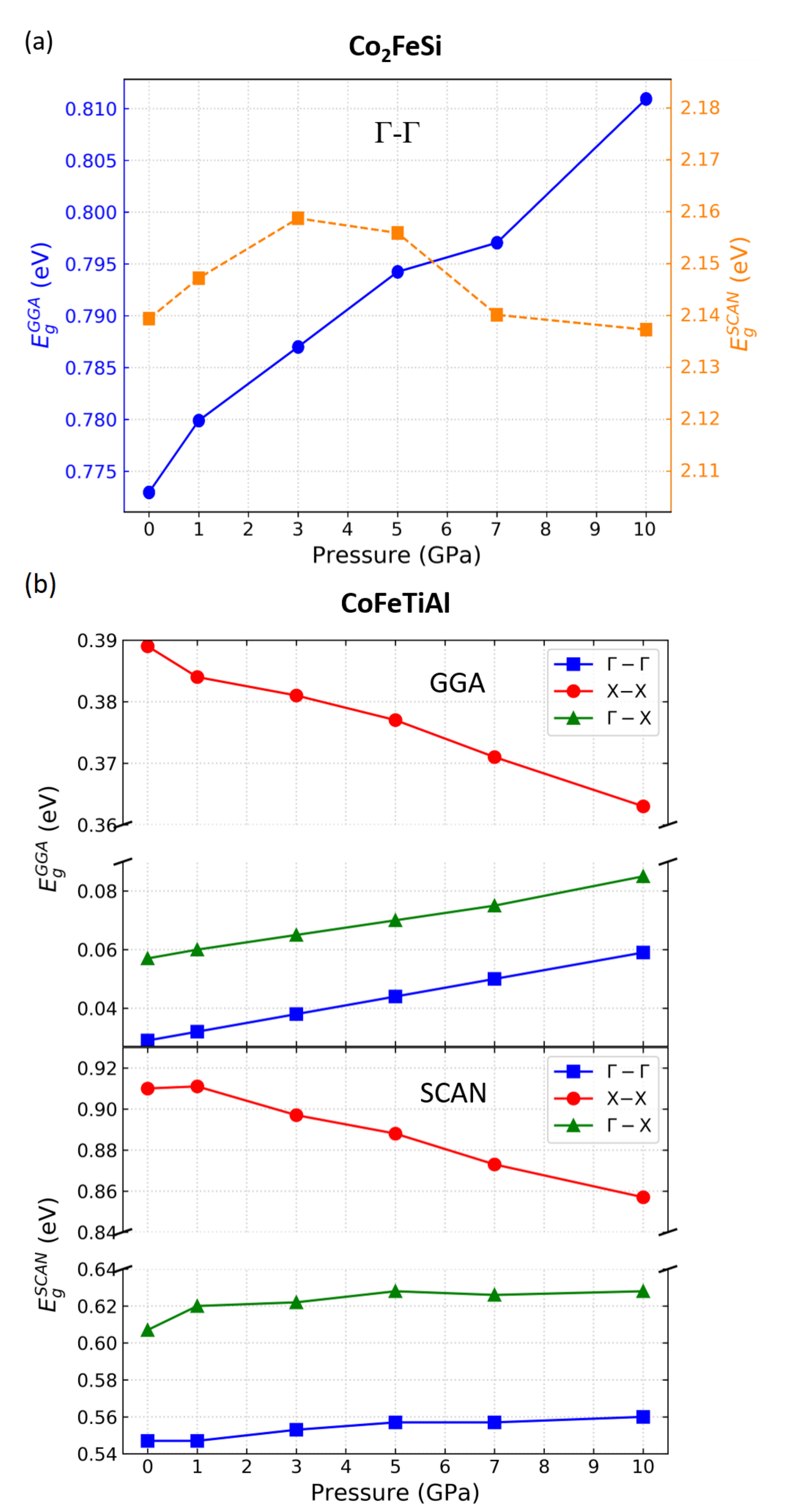}}
    \caption{ Transition energies between high symmetry points in dependence of external pressure for (a)~Co$_2$FeSi and (b)~CoFeTiAl calculated with GGA and SCAN.}
    \label{fig3}
\end{figure}

Next, we consider the effect of external pressure on the structural, magnetic, and electronic properties.
We find that GGA yields a linear decrease in the lattice parameter with an applied pressure with a slope 0.008~\AA/GPa for both Co$_2$FeSi and CoFeTiAl.
However, SCAN results are slightly different for ternary and quaternary alloys, since the slope is 0.008~\AA/GPa for Co$_2$FeSi and 0.009~\AA/GPa for CoFeTiAl.
The magnetic moment of Co$_2$FeSi decreases linearly with pressure for both GGA and SCAN.
However, the SCAN slope is 0.009~$\mu_B$/GPa, which is almost two times less than the GGA slope given by 0.017~$\mu_B$/GPa. 

Concerning the electronic properties, the influence of external pressure on the transition energies between high-symmetry points is illustrated in Fig.~\ref{fig3}.
Energy gaps are affected by pressure since unoccupied and occupied states are pushed up and down, respectively, under the influence of external pressure. 
The $\Gamma-\Gamma$ and $\Gamma-X$ energy gaps  increase linearly  while the $X-X$ energy gap decreases with pressure for both alloys. 
The energy gap widening at $\Gamma$ point for Co$_2$FeSi and CoTiFeAl produces insulating characteristics. These trends are predicted both by GGA and SCAN. However, the SCAN gap values are an order of magnitude higher than those obtained by GGA.

\section{Conclusions}
We have considered corrections beyond GGA for structural, magnetic, and electronic properties of Co$_2$FeSi and CoTiFeAl full-Heusler compounds.
SCAN gives a magnetic moment for Co$_2$FeSi consistent both with the Slater-Pauling rule and with the experimental value.
Moreover, the half-metallic behavior of Co$_2$FeSi and semiconductor behavior of CoTiFeAl predicted by SCAN are in good agreement with experiments.
We conclude that SCAN can capture similar corrections to those included in the GGA$+U$ and $GW$ schemes while avoiding external $U$ parameters and overwhelming computational costs.

\begin{acknowledgements}
This work was supported by Russian Science Foundation~No.~17-72-20022. 
O.M. and M.Z. acknowledge support from the Young Scientist Support Fund of Chelyabinsk State University.
B.B. acknowledges support from the COST Action CA16218.

\end{acknowledgements}

\section*{AIP Publishing Data Sharing Policy}
The data that support the findings of this study are available from the corresponding author upon reasonable request.

\label{References}

\bibliography{RevisedBIB}

\begin{thebibliography}{29}
\bibitem [{Sla()}]{Slater-Pauling}%
  \BibitemOpen
  \href@noop {} {}\bibinfo {note} {Slater-Pauling rule connects the number of
  valence electrons $N_V$ with the value of $\mu_{tot}$/f.u. as
  $\mu_{tot}\mathrm{/f.u.}=N_V-24$. Co$_2$FeSi should have the integer-value of
  magnetic moment since its $N_V=30$.}\BibitemShut {Stop}%
\end{thebibliography}
\end{document}